\documentclass[11pt,a4paper]{article}
\usepackage{cite}
\usepackage{amsmath, amsthm, amssymb,slashed}
\usepackage{yfonts}

\usepackage{ifpdf}
\ifpdf
  \usepackage[pdftex]{graphicx}
  \usepackage{epstopdf}
\else
  \usepackage[dvips]{graphicx}
\fi

\usepackage{subfigure}
\usepackage{xfrac}  
\usepackage{fancyhdr}
\usepackage{mathbbol}
\usepackage{url}
\usepackage{color}
\usepackage{bbm}
\usepackage{rsfso}
\usepackage{dsfont}
\usepackage{bm}
\usepackage{amsfonts}
\usepackage{mathrsfs}
\usepackage{slashed}
\usepackage{wasysym}
\usepackage{yfonts}
\usepackage{ytableau}
\usepackage{tikz}
\usetikzlibrary{arrows,decorations.pathmorphing,hobby}
\usetikzlibrary{calc,arrows.meta}
\usepackage{cancel}
\usepackage[normalem]{ulem}%provides strikeout through \sout{ }

\def\bl{\begin{equation}\begin{aligned}}
\def\el{\end{aligned}\end{equation}}
\def\beal{\begin{align}}
\def\eal{\end{align}}
\def\be{\begin{equation}}
\def\ee{\end{equation}}
\def\bpm{\begin{pmatrix}}
\def\epm{\end{pmatrix}}
\def\bsm{\begin{bmatrix}}
\def\esm{\end{bmatrix}}
\def\bvm{\begin{vmatrix}}
\def\evm{\end{vmatrix}}
\def\bVM{\begin{Vmatrix}}
\def\eVM{\end{Vmatrix}}
\def\bea{\begin{eqnarray}}
\def\eea{\end{eqnarray}}

\def\1{{\bf 1}}
\def\2{{\bf 2}}
\def\3{{\bf 3}}
\def\4{{\bf 4}}

\usepackage{indentfirst}

\newcommand{\OMIT}[1]{}

\usepackage{mathtools}

\title{Radiative decays of $X(3872)$ discriminate between the molecular and compact interpretations. }
\date{\today}
\author{B. Grinstein$^{\P}$, L. Maiani$^*$ and A.D. Polosa$^*$\\{\it $^{\P}$University of California at San Diego, 9500 Gilman Drive, La Jolla, CA 92093, USA} \\ {\it $^*$Sapienza University of Rome and INFN, Piazzale Aldo Moro 2, I-00185, Italy}   }
\begin{document}
\maketitle
%\begin{document}

\begin{abstract}
%   Qualche nota $\mcal{ABCPFT}$  and $\scal{A}{B}{C}$ and $\expv{A}$ and $\ketbra{A}{B}$
%   $\mathfrak{ABCDEFGHILMNOPQRSTUVWXYZ}$ and $\mathfrak{abcdefghijklmnopqrstuvwxyz}$
%\textfrak{This is frak} \textswab{This: is: Schwabacher} \textgoth{This: is: Gothic}
%{\frakfamily\fraklines
%   \yinipar{T}his: is: an initial with a sentence in Fraktur.\\
%   \textfrak{This: is: also Fraktur.}\\
%   \textswab{This: is: Schwabacher.}\\
%   \textgoth{This: is: Gothic.}
%}
%{\frakfamily\fraklines
%   \yinipar{R}
%}
Radiative decays $X\to \psi(1S) \, \gamma$ and $X\to \psi^\prime(2S)\, \gamma $ might be expected to have a ratio of branching fractions following  the  phase space volumes ratio.  However data suggest the opposite, indicating a value for  ${\cal R}={\cal B}(\psi^\prime \gamma)/{\cal B}(\psi \gamma)$ consistently larger than one. In this paper we present a calculation of ${\cal R}$ for both a compact Born-Oppenheimer $c\bar c q\bar q$ state and a $D\bar D^{*}$ molecule.  In the former case ${\cal R}\sim 1$  or  larger is found, a value to be confronted with forthcoming high statistics data analyses. In the molecular picture, with $D$ and $\bar D^{*}$ mesons described by the universal wave function used by Voloshin, Braaten and Kusunoki, we find that ${\cal R}$ would be of order $10^{-2}$. A more precise experimental measure  would be extremely helpful in clarifying  the true nature of the $X(3872)$.

\end{abstract}

\section{Introduction}
%\forceindent
It is widely accepted that the $X(3872)$ is a tetra-quark. There are
two competing models for the way in which the two quarks, $c$ and $u$,
and the two anti-quarks, $\bar c$ and $\bar u$ are distributed in the
$X$.\footnote{We ignore the
  fact that the $X(3872)$ may be a superposition
  $\alpha [c\bar c u\bar u]+\beta[c \bar c d\bar d]$, since the
  dependence from $\alpha$ and $\beta$ cancels in the ratio of the
  decay rates into $\psi^{(\prime)}~\gamma$.} 
  In the ``molecular" model a weakly bound state is formed of
a $D$ and a $\bar D^*$ mesons. This molecule is very big. The reason
is as follows.  The attractive force responsible for the binding is
described by a spherical potential well, much like the nuclear force
that binds nucleons in a nucleus. The known mass of the $X$ implies an
extremely small binding energy resulting in a very large
wave-function. The picture in the alternative ``compact" tetra-quark
model is quite different: here a $uc$ pair binds into a color
anti-triplet, which makes a bound state via the Coulomb-plus-linear
potential with the $\bar c\bar u$ color triplet. This is a different
kind of object: the overall size is significantly smaller than the
molecular model's, and the pairs of quarks are bound into colored
objects that are significantly larger than the charmed mesons of
the molecular model.

The $X(3872)$ is certainly the outlier with respect to all other
exotic resonances observed so far in that it has a mass almost
perfectly equal to the sum of the masses of $D$ and $D^*$ mesons. This
represents a peculiar source of fine tuning in both of the above
interpretations~\cite{Esposito:2023mxw}.

The meson constituents of the loosely bound  $X$ state are expected to have a relative 
momentum in the center of mass\footnote{This can be computed assuming a $\lambda \delta^{3}(\bm r)$ potential binding the $D\bar D^*$ pair and  using the quantum virial theorem. }  $p\lesssim 30$~MeV and there are very few $D\bar D^*$ pairs in that kinematical region in high energy collisions, especially when   high $p_T$ cuts are included\cite{Bignamini:2009sk,Artoisenet:2009wk,Esposito:2015fsa}. 
On the other hand a compact component of the $X$ in the description of the prompt production in $pp$ collisions allows to explain  the high production cross sections observed.
Problems with the production of the $X$ are found also in high multiplicity final states. In~\cite{Esposito:2020ywk} it is shown that the prediction for the $X$ yield in high multiplicity final states at LHCb are in contrast with data, if a coalescence model  is used. The same model works very well with deuteron, the loosely bound state which is often compared with the $X$.

Then there is the much discussed problem of the determination of the
effective radius $r_0$ from the line-shape of the $X(3872)$ made by
the LHCb collaboration~\cite{LHCb:2020xds}. According to
Weinberg~\cite{Weinberg:1965zz}, a negative and large effective radius
can be taken as the token of a compact particle, if in combination
with a positive scattering length, and this is the case of the
measured $r_0$, as discussed in~\cite{Esposito:2021vhu}. The
conclusions reached in~\cite{Esposito:2021vhu} are corroborated by the
analysis in~\cite{Esposito:2023mxw}~and, albeit with larger errors, by
the BESII collaboration;~\cite{BESIII:2023hml}\footnote{In the
  convention where $ f^{-1}=-1/a +1/2 r_0 k^2 $,
  Ref.~\cite{Esposito:2021vhu}, based on LHCb
  data~\cite{LHCb:2020xds}, and BESIII~\cite{BESIII:2023hml} give
%\beal
\begin{align}
 a&\simeq +28~\text{fm} && -2.0 \lesssim  r_0 \lesssim -5.3~\text{fm} & (\text{LHCb}) \notag\\ %~\cite{Esposito:2021vhu} \notag \\
a&=+16.5^{+27.6+27.7}_{-7.0-5.6}~\text{fm} &&  ~r_0=-4.1^{+0.9+2.8}_{-3.3-4.4}~\text{fm}&   (\text{BESIII})\notag %~\cite{BESIII:2023hml}
\end{align}
%\eal
In a very recent paper it is claimed that a combined analysis of LHCb and Belle data gives $r_0\simeq-14$~fm\cite{Xu:2023lll}. 
}  see also~\cite{Braaten:2020nmc}.

In this  paper we point out that also $X(3872)$
radiative decays can give a strong indication about the nature of the
$X$. Indeed the ratio of branching ratios
\begin{equation}
    {\cal R}=\frac{{\cal{B}} (X\to\psi'\gamma)}{{\cal{B}}(X\to \psi\gamma)}
    \label{eq:RatioDefd}
  \end{equation}
  observed in data, being of order unity or larger, is in conflict
  with basic molecular models, unless specific assumptions on the
  couplings are made~\cite{Guo:2014taa}. In the following we present a
  calculation of this ratio in both scenarios, finding an ${\cal R}$
  value of order one for the compact tetraquark, about thirty times
  larger than the expected ${\cal R}$ found with minimal molecular
  assumptions.

  The precision in the measurement of this ratio is dramatically
  improving recently and we look forward to the forthcoming more
  precise determination of ${\cal R}$. The value from the
  PDG\cite{ParticleDataGroup:2022pth} is approximately
  ${\cal R}\simeq 6\pm 4$. The reason why ${\cal R}$ should
  discriminate well between models is that the final state charmonium
  has much larger spatial extent in the numerator $\psi'$ than in the
  denominator $J/\psi$, and in order to produce a photon via $u\bar u$
  annihilation the two quarks have to come to a common point. This
  will be described in the next section.

\section{Modeling $X$ radiative decay}
\label{Sec:famework}
% \forceinden

Assuming the tetraquark has no significant charmonium component, so that it  is truly a {\it tetra}-quark, its radiative decay must involve $u\bar u$ annihilation. To lowest order this  is from  $u\bar u \to \gamma$. We adopt a non-relativistic potential quark model for the tetraquark. The 4-quark wavefunction  contains fast ($u$-quarks) and slow ($c$-quarks) degrees of freedom, and much like in molecular physics the full wavefunction can be well approximated using the method of Born  and Oppenheimer~\cite{Maiani:2019cwl,wqm2}. 

The full wavefunction $\Psi({\bf r}_c,{\bf r}_{\bar c},{\bf r}_u,{\bf r}_{\bar u})$ 
is approximated by the product of wavefunctions of fast and slow degrees of freedom. The former is computed as the wavefunction of the $u$-quarks in the potential due to static sources of color charge produced by the $c$-quarks. Moreover, in our work we will  approximate this  as the product of separate ``atomic'' wave-functions, $\chi_{_\text{M}}(|{\bf r}_{u}-{\bf r}_{\bar c}|)\chi_{_\text{M}}(|{\bf r}_{\bar u}-{\bf r}_c|)$  for the molecular picture 
and $\chi_{_\text{C}}(|{\bf r}_{u}-{\bf r}_{ c}|)\chi_{_\text{C}}(|{\bf r}_{\bar u}-{\bf r}_{\bar c}|)$ 
for the compact tetraquark. 

These are used to compute the energy of the system as a function of separation between the $c$ and $\bar c$, which is used as a potential in the computation of the wavefunction $\Psi_{{\rm BO}}(|{\bf r}_{c}-{\bf r}_{\bar c}|)$ 
of the $c$ and $\bar c$ 2-body system. Thus we have
$\Psi({\bf r}_c,{\bf r}_{\bar c},{\bf r}_u,{\bf r}_{\bar u})\approx
\chi_{_\text{C}}(|{\bf r}_{u}-{\bf r}_{ c}|)\chi_{_\text{C}}(|{\bf
  r}_{\bar u}-{\bf r}_{\bar c}|)\Psi_{{\rm BO}}(|{\bf r}_{c}-{\bf r}_{\bar
  c}|) $
  in the compact tetraquark picture and
 $\Psi({\bf
  r}_c,{\bf r}_{\bar c},{\bf r}_u,{\bf r}_{\bar u})\approx
\chi_{_\text{M}}(|{\bf r}_{u}-{\bf r}_{\bar c}|)\chi_{_\text{M}}(|{\bf
  r}_{\bar u}-{\bf r}_c|) \Psi_{\rm mol.}(|{\bf r}_{c}-{\bf r}_{\bar
  c}|) $  in the molecular picture. The $\Psi_{\rm mol.}$ wavefunction 
  derives from the treatment of shallow bound states in non-relativistic 
  scattering theory, see Section~\ref{moles}.
  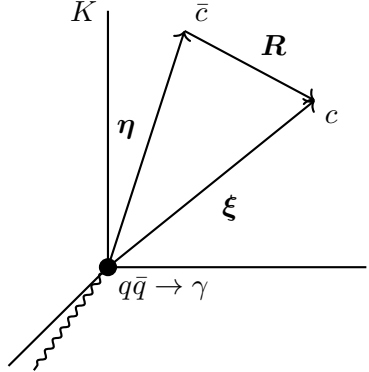
\begin{figure}[ht!]
  \centering
  \begin{tikzpicture}[scale=1.7]

    % Draw the axes
%    \draw[thick,->] (0,0,0) -- (2,0,0) node[below]{$y$};
%    \draw[thick,->] (0,0,0) -- (0,2,0) node[left]{$z$};
%    \draw[thick,->] (0,0,0) -- (0,0,2) node[below]{$x$};
    \draw[thick] (0,0,0) -- (2,0,0);
    \draw[thick] (0,0,0) -- (0,2,0) node[left]{$K$};
    \draw[thick] (0,0,0) -- (0,0,2);

    % Define the coordinates of point C and P
    %\coordinate (C) at (1,1,0);
    \coordinate (C) at (1.6,1.3,0);
    \coordinate (P) at (4.25,5.5,9.5);

    % Draw vectors from the origin to C and P
    \draw[->, thick] (0,0,0) -- (C) node[midway, anchor=north west]{$\bm \xi$};
    \draw[->,thick] (0,0,0) -- (P) node[midway, anchor=south east]{$\bm \eta$};

    % Draw a line connecting C and P
    \draw[->, thick] (P) -- (C) node[midway, anchor=south  west]{$\bm R$};

    % Label the points
    \node[below right] at (C) {$c$};
    \node[above right] at (P) {$\bar c$};
    
     % Add a small disk (dot) at the origin
    \fill (0,0,0) circle (2pt);
      \coordinate (O) at (0,0,0);
     \node[below right] at (O) {$q\bar q\to \gamma $};
    
      \draw[thick, decorate, decoration={snake, amplitude=1pt, segment length=5pt}] (0,0,0) -- (0.05,-0.18,1.62);

  \end{tikzpicture}
  \caption{\footnotesize{The light quarks annihilate in the origin of the frame $K$, where the $X$ is initially at rest. The photon is taken in the $xz$ plane.}}
\end{figure}

In this non-relativistic setting the calculation proceeds identically
for the molecular and compact models. The distinction between these is
exclusively from the different wavefunctions
adopted.\footnote{Additional distinctions from, {\it e.g.}, color factors cancel in the ratio ${{\cal R}}$.}

Without loss of generality, we assume that  the  annihilation takes
place in the origin of the tetraquark’s rest-frame $K$ in Fig.~1.  Let
$\psi$ be the wavefunction of the $\psi(1S)$ or  $\psi^\prime (2S)$. 
The transition amplitude $A$ in the $X$ rest frame, at fixed photon three-momentum $\bm k$, is\footnote{The integrals in
Eq.~\eqref{main} may be computed by choosing frame orientations such that
\be
\bm
\xi =
\bpm
r \sin\theta\cos\phi\\
r\sin\theta\sin\phi\\
r\cos\theta
\epm
\qquad
\bm R =
\bpm
R\sin\lambda\\
0\\
R\cos\lambda
\epm
\qquad
\bm k=
\bpm
 0\\0\\k
\epm 
\ee
so that $d^3\xi =r^2dr\,d\cos\theta\,d\phi$ and $d^3R=2\pi R^2\,
dR\:d\cos\lambda$. Here
  $k=(M_{X}^2-M^2_{\psi^{(\prime)}})/2M_X $.}
\begin{align}
A (X\to \Psi\, \gamma)
&={\cal F}\int
d^3R\,d^3\xi\,d^3\eta\,\;\delta^3(\bm\eta+\bm R-\bm \xi) e^{-i
  \frac12\bm k\cdot (\bm \xi +\bm \eta)}\,  {\psi (|\bm R|)}\, \Psi(|\bm R|)\, \chi(|\bm \xi|)\,
                               \chi(|\bm \eta|) \nonumber\\
  \label{main}
  &={\cal F}\int
d^3R\,d^3\xi\; e^{-i \bm k\cdot \left(\bm \xi -\frac{\bm R}{2} \right)}\, \psi (|\bm R|)\, \Psi(|\bm R|)\, \chi(|\bm \xi|)\, \chi(|\bm \xi-\bm R|)
\end{align}
where $\chi$ can be either  $\chi_{_{\rm M}}$ or $\chi_{_{\rm C}}$ and $\Psi$ can be $\Psi_{\rm mol.}$ or $\Psi_{\rm BO}$ respectively, whereas $\psi$ is the charmonium wavefunction.  
The  exponential factor takes into account the recoil of the $c\bar c$
  pair against the photon emitted in the  $q\bar q$ annihilation.\footnote{Upon photon emission the heavy quarks recoil against $k$, the momentum of the photon, with  velocity $v\sim  k/2M_c$. This allows to use a Galileo boost $\exp(-i\bm K\cdot \bm  v)$ on the quantum state  $\Phi$ of the heavy quarks. Since   $\exp(i\bm K \cdot \bm v)\Phi_{\bm q}=\Phi_{\bm q-2M\bm v}$ on momentum eigenstates,  the boost  introduces a phase in the wavefunction $\psi(\bm x)=(\Phi_{\bm x},\Phi)$, equal to $\exp(i\frac12\bm k\cdot(\bm \xi+\bm \eta) )$. This  gives the phase  used in~\eqref{main}, where $\psi^*(\bm x)$ is taken.  }  
All factors which get cancelled in the ratio of branching ratios
${\cal R}={\cal B}(\psi^\prime \gamma)/{\cal B}(\psi \gamma)$ are
absorbed in the fudge factor ${\cal F}$, except for the product of the
polarization vectors of $X,\psi^{(\prime)}$ and $\gamma$ which comes
in the combination of a mixed product $ \epsilon(\bm
e^*_{(\psi^{(\prime)})},\, \bm e^*_{(\gamma)},\, \bm
e_{(X)})$. Summing its square modulus over polarizations in the rest
frame of the $X$ 
  \be
{\cal S}_{\psi^{(\prime)}}=\sum_{\rm pols} |\epsilon(\bm e^*_{(\psi^{(\prime)})},\bm e^*_{(\gamma)},\bm e_{(X)})|^2=\epsilon_{ijk}\epsilon_{i^\prime j^\prime k}
(\delta_{ii^\prime}+\frac{k_ik_{i^\prime}}{M_{\psi^{(\prime)}}^2})(\delta_{jj^\prime}-\frac{k_j k_{j^\prime}}{\bm k^2})=4+2\frac{\bm k^2}{M^2_{\psi^{(\prime)}}}
\ee
where 
\(
|\bm k|=\frac{(M_X^2-M^2_{\psi^{(\prime)}})}{2 M_X },
\)
one finds ${\cal S}_{\psi^\prime}/{\cal S}_{\psi}=0.98$.

Only the real part of
the exponential factor, 
contributes appreciably to the amplitude and all the plots in the following are calculated using  the real part. 

For the charmonium wave function {$\psi (R)$}, we solve numerically  the Schr\"odinger
  equation in the Cornell potential\cite{Cornell}.
\be
V(R)=-\frac{4}{3}\frac{\alpha_s}{R}+{\kappa}\,  R
\label{stcornell}
\ee
with $\alpha_s=0.331$ and $\kappa=0.18$~GeV$^{2}$ and $M_c=1.317$~GeV for the charm quark mass~\cite{Soni:2017wvy}. With these parameters we
reproduce well the experimental difference in mass between $\psi(2S)$
and $\psi(1S)$.

\begin{figure}[ht!]
\centering
 \includegraphics[width=0.6\textwidth, angle=0]{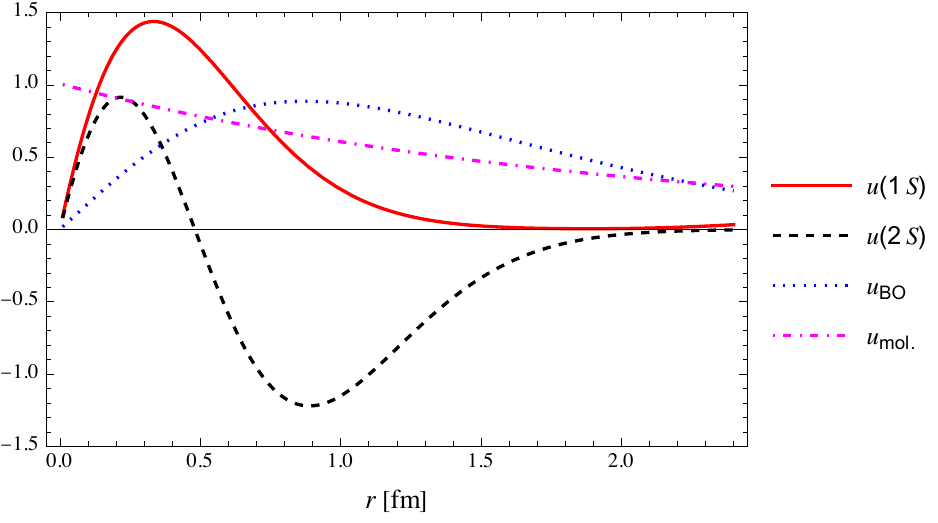}
 \caption{\label{reduced} \footnotesize{The reduced, normalized, wavefunctions
   $u(r)=r R(r)$, where $R(r)$ is the radial component of the
   wavefunction, $\Psi(r)=R(r)\, Y_{\ell m}$.   For standard
   $1S$ and $2S$ charmonia, the reduced normalized functions are $u(1S)$ and $u(2S)$, and for the
   $c\bar c$ component in the  universal molecular wavefunction discussed in
   Section~\ref{moles}, we plot the normalized $u_{\rm mol.}$ in place of $\Psi_{\rm mol.}$.  
   As for  the compact  tetraquark
   computed in the Born-Oppenheimer potential  in
   Eq.~\eqref{eq:BOpot},  the reduced wavefunction is $u_{\rm BO}$. }}
\end{figure}

 For the $1S$ and $2S$ charmonium we use  the  ground state
and first excited state eigenfunctions, respectively, Fig.~\ref{reduced}.

\section{Compact Tetraquark radiative decay}
\label{compa}
The diquark wavefunction $\chi_{_{_\text{C}}}$ is evaluated using the variational
principle. The two body hamiltonian $H=K+V$ is assumed to have
potential%
\be
V(r)=-\frac{1}{3}\frac{\alpha_s}{r}+{\kappa^\prime}\,r\,,
\label{stension}
\ee

The smallest diquarks are obtained with $\kappa^\prime=\kappa$, larger sizes can be obtained by decreasing $\kappa^\prime$  down to $\kappa^\prime=\frac14 \kappa$ as predicted in the Born-Oppenheimer picture we use.\footnote{\label{scalcas}We consider the color arrangement
  $|(c\bar c)_{\bm 8}(q\bar q)_{\bm
    8}\rangle=\sqrt{\frac{2}{3}}|(cq)_{\bar{\bm 3}}(\bar c\bar q)_{\bm
    3}\rangle- \sqrt{\frac{1}{3}}|(cq)_{\bar{\bm 6}}(\bar c\bar
  q)_{\bm 6}\rangle$ so that the coupling
  $\lambda_{cq}=\frac{2}{3}\frac{1}{2}(-\frac{4}{3})+
  \frac{1}{3}\frac{1}{2}(\frac{2}{3})=-\frac{1}{3}$. In~\cite{Bali:2000gf}
  it is discussed that the string tension in the potential formula
  should also scale with the $\lambda$ coupling (Casimir scaling). In
  the formula~\eqref{stcornell} for the Cornell potential we would
  confirm
  $\kappa_{\rm eff.}=\frac{3}{4}|\lambda_{c\bar c}|\kappa
  =\frac{3}{4}|-\frac{4}{3}|\kappa=\kappa$. But in place of $\kappa$
  in~\eqref{stension} we might also consider a smaller effective string
  tension, making a larger diquark:
  $\kappa_{\rm eff.}=\frac{3}{4}|\lambda_{cq}|\kappa
  =\frac{3}{4}|-\frac{1}{3}|\kappa=\frac{1}{4}\kappa$.  }

A trial wavefunction
\be 
  \chi_{_\text{C}}(r)=\frac{2{\cal C}^{3/2}}{\sqrt{4\pi}}e^{-{\cal C}
        r}\label{orbitalwf}
 \ee
    is used, with the constant ${\cal C}$ determined by
    minimizing $(\chi, H\, \chi)/(\chi,\chi)$.
In the calculation of the amplitude in Eq.~\eqref{main}, the Born-Oppenheimer wavefunction is used
\be
\Psi(R)\to \Psi_{\rm BO}(R)
\ee
as computed from the potential 
\be
\label{eq:BOpot}
V_{\rm BO}(R)=\frac{1}{6} \frac{\alpha_s}{R}- 2 \frac{7\alpha_s}{6} I_1(R)+ \frac{\alpha_s}{6} I_4(R)  + k\, \theta(R-R_{0s})\,(R-R_{0s})
 \ee
 %
% where {\color{orange} $R_{0s} $ is the starting point of the string that confines the two, color triplet, orbitals~\cite{Maiani:2019cwl}}.  
The first term in the
 BO potential corresponds to the octet repulsion between the two
 heavy quarks. The term containing the function $I_1(R)$ corresponds
 to $c\bar q$ and $q\bar c$ interactions, with the Fierz coefficient
 7/6 calculated as detailed in\cite{Maiani:2019cwl}.  The term
 containing the $I_4(R)$ function describes $q\bar q$ interactions,
 and the same octet coefficient of $1/6$ appearing in the first term
 is included here. The functions $I_1$ and $I_4$ are given by
\bea
&&I_1(R)=\int_{\bm \xi }{\chi_{_\text{C}}(|\bm \xi|)^2}\frac{1}{|\bm \xi-\bm R|} \notag\\%=\int_{\bm \eta}\phi(\bm \eta)^2\frac{1}{|\bm \eta+\bm R|} ,\notag \\
&&I_4(R)=\int_{\bm \xi,\bm \eta }{\chi_{_\text{C}}(|\bm \xi|)^2} \chi_{_\text{C}}(|\bm \eta|)^2 \frac{1}{|\bm \xi-\bm R-\bm \eta|} 
\eea
Notice that the functions $I_1(R)$ and $I_4(R)$ are computed in the
generic case in which the light quarks are not found at a single
point, like in the origin of Fig.~1. Finally, we include in $V_{\rm BO}$
a confining, linearly rising, potential of the
colored diquarks, starting at $R=R_{0s}$. %we take to become significant at a hadron sizeof approximately $R_{0s}= 3\pm 1$~fm
For orientation, we choose
$R_{0s}=3\pm 1$~fm, i.e. greater than $2{\cal C}^{-1}$ which is the size where the two
orbitals start to separate.

\begin{figure}[ht!]
\centering
\includegraphics[width=0.45\textwidth, angle=0]{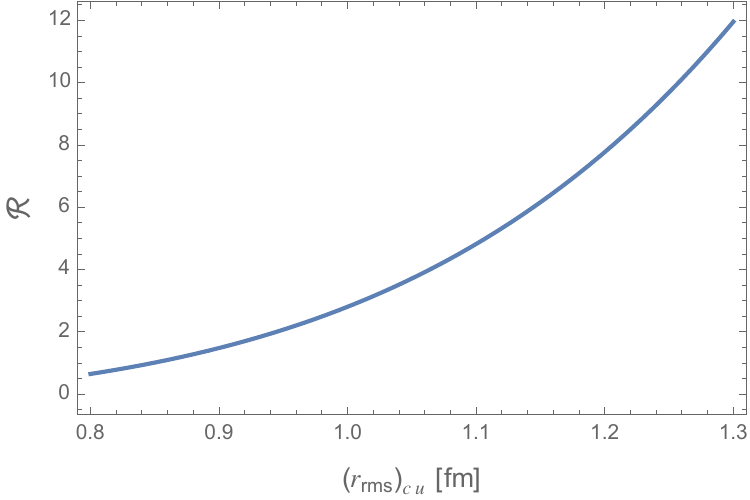}
\caption{\footnotesize{The ratio ${\cal R}$ as a function of the size $(r_{\text{rms}})_{cu}$ of the diquarks in the compact tetraquark picture. For the two values discussed in the text after Eq. (11), $ {\cal R}$ varies from $ {\cal R}\sim1$, at $(r_{\text{rms}})_{cu}= 0.83$ fm to $ {\cal R}\sim12$, at $(r_{\text{rms}})_{cu}  = 1.3$ fm. 
}}
%The ratio ${\cal R}$ as a function of the size $r_{\rm rms}$ of the diquarks in the compact tetraquark picture,{ \color{blue} including the two values  $(r_{\rm rms})_{cu}=0.83$ and  $(r_{\rm rms})_{cu}$ discussed in the text}. The attraction in the diquark is weaker than that in a singlet meson, thus we might expect a larger size for the diquark than for the $D/D^*$  mesons.\\
  % \BG{(i) I do not understand what the sentence in blue is meant for, nor what the ``and   $(r_{\rm rms})_{cu}$'' with no value means, and \\  (ii) For notational consistency we should have the x-axis as ``$(r_{\rm rms})_{cu}$~[fm]'' instead of ``Size of diquark $\sqrt{\langle r^2\rangle_{\psi_c}}$ (in fm)''. Similar changes should be done in Figs 4 and 5 as well.  
   \label{sizedq}
\end{figure}

With the given parameters we find that  the smallest possible value for ${\cal R}$ in the compact picture is 
\be
\label{err0s}
{\cal R}_{\rm min} = \frac{{\cal B}(X\to \psi^\prime \gamma)}{{\cal B}(X\to \psi  \gamma)}
=0.95^{+0.01}_{-0.07}
\ee
despite the fact that the ratio of the phase space volumes is
$\Phi(\psi^\prime \gamma)/\Phi(\psi \gamma)=0.26$. The reason for this
can partially be captured by comparing the $\Psi_{\rm BO}(R)$ with the
charmonium $\Psi(R)$, as done in Fig.~2. Note, however, that the
superposition integral~\eqref{main} includes as well the oscillating
factor %~\eqref{ofa} 
 which is not displayed in Fig. 2. We will return to
this point in the discussion Section~\ref{discussion1}. 
For the moment observe that the value ${\cal R}_{\rm min}$ is obtained using $\kappa^\prime =\kappa$ in~\eqref{stension} which makes the smaller diquarks size $(r_{\rm rms})_{cu}\simeq 0.83$~fm, see Fig.~\ref{sizedq}. Allowing smaller string tensions $\kappa^\prime$ makes looser diquarks, as large as $(r_{\rm rms})_{cu}=1.3$~fm in size, corresponding to $\kappa^\prime =\frac14 k$, see Fig.~\ref{sizedq}.

\section{Molecule radiative decay\label{moles}}
There is a universal prediction for the $D^0 \bar D^{*0}$ or $\bar D^0 D^{*0}$ wavefunction~\cite{Voloshin:2003nt, Braaten:2003he} (see also~\cite{Jackiw:1991je} and especially the discussion in~\cite{lqm3-2})
\be
\Psi_{\rm mol.}(R)=\frac{1}{\sqrt{2\pi R_0}}\frac{e^{-R/R_0}}{R}
\label{mol}
\ee
where $R_0\simeq 1/\sqrt{2m B}$, $B$ being the molecule binding energy. We assume  $R_0\approx 10$~fm ($B\simeq 200$~KeV)
and in the calculation of the amplitude in~\eqref{main} we substitute 
\be
\Psi(R)\to \Psi_{\rm mol.}(R)
\ee
as given in~\eqref{mol}. 
\begin{figure}[ht!]
\centering
 \includegraphics[width=0.5\textwidth, angle=0]{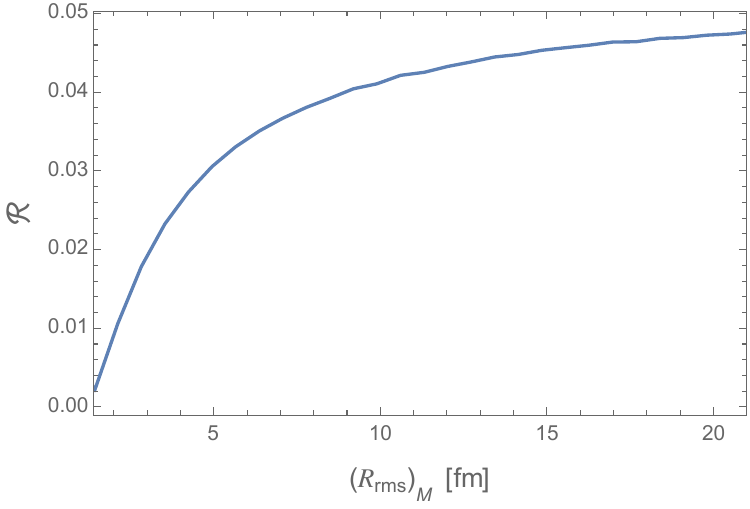}
 \caption{\footnotesize{The ratio ${\cal R}$ as a function of the molecular size  around the value $(R_{\text{rms}})_{_\text{M}}=7$ fm ($R_0=10$~fm ) discussed in the text}.}
 \label{Fig:MolSize}
\end{figure}

{For the quark orbitals in the $D$ and $D^*$ mesons, $\chi_{_{\rm {M}}}$, we
  use}  Isgur-Scora-Grinstein-Wise (ISGW) functions~\cite{Isgur:1988gb}
\be
\label{isgwe}
{\chi_{_{\rm M}}(r)}=\frac{b^{3/2}}{\pi^{3/4}}e^{-\frac{1}{2}b^2 r^2}
\ee
where $b=0.35$~GeV  (giving $(r_{\rm rms})_{D,D^*}\simeq0.69\;\text{fm}$) is calculated at the given value of $\alpha_s$.

Using $(R_{\rm rms})_{_\text{M}}=17\;\text{fm}$ and $(r_{\rm rms})_{D,D^*}=0.68\;\text{fm}$ we find 
\be
{\cal R}=\frac{{\cal B}(X\to \psi^\prime \gamma)}{{\cal B}(X\to \psi \gamma)}=0.036\,.
\ee
As can be seen in Fig.~\ref{Fig:MolSize}, this ratio slowly saturates at larger molecular sizes remaining quite smaller than the observed value. 
The ratio ${\cal R}$ found {for}  the compact tetraquarks, Eq.~\eqref{err0s}, is therefore
at least  thirty times larger than that of the molecular picture.

\begin{figure}
    \centering
    \begin{minipage}{0.45\textwidth}
        \centering
        \includegraphics[width=0.9\textwidth]{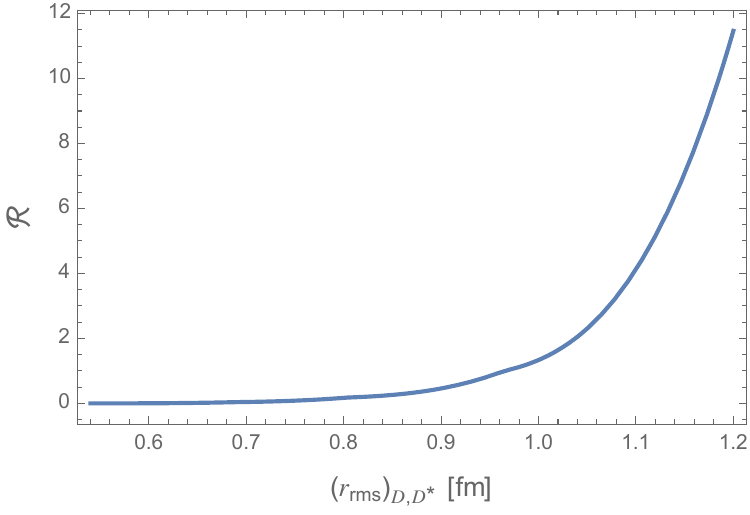} % first figure itself
        %\caption{first figure}
    \end{minipage}\hfill
    \begin{minipage}{0.45\textwidth}
        \centering
        \includegraphics[width=0.9\textwidth]{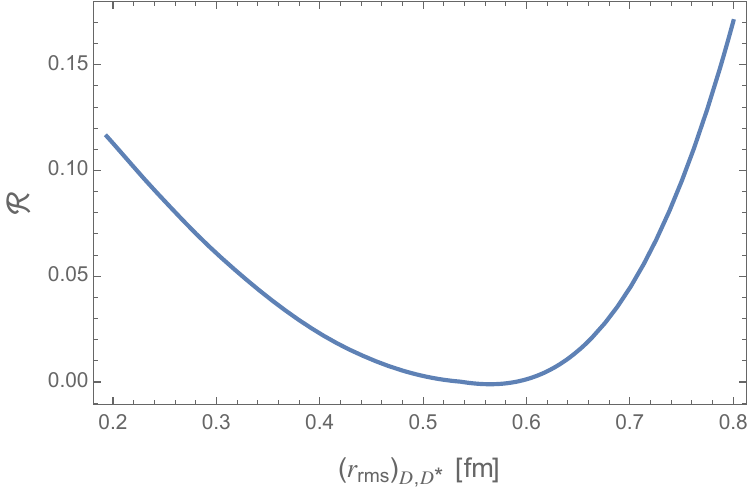} % second figure itself
        %\caption{second figure}
    \end{minipage}
    \caption{\footnotesize{The ratio ${\cal R}$ as a
      function of the size of the $D$ or $D^*$ meson keeping the
      characteristic size of the molecule $R_{\rm rms}\simeq 7$~fm. The value
      used in the text is
      $(r_{\rm rms})_{D,D^*}\equiv\sqrt{\langle r^2\rangle_{\psi_{_{\rm M}}}}\simeq
      0.69$~fm. With respect to Fig.~\ref{sizedq}, the small distance
      region has a much lower ${\cal R}$ and is flatter. }}
      \label{bgmol} 
\end{figure}

\section{Discussion}
\label{discussion1}
The calculation above shows a %tremendous 
remarkable disparity in the predictions
of the $2S$-to-$1S$ ratio of radiative branching fractions $\mathcal{R}$.
% These predictions are made within the models delineated above.
Working within the framework explained in Sec.~\ref{Sec:famework}, but
artificially modifying  the wavefunctions,   we
can investigate the dependence of our results on the specific
assumptions of the models used for the compact and molecular pictures
of the tetraquark. We will see that the main conclusion, that
$\mathcal{R}$ is much bigger for the compact tetraquark than for the
molecular picture, is very robust. The detailed shape of the
wavefunctions  matter little. The determining factor turns out to be
the singular nature of the universal wavefunction in the molecular picture, \eqref{mol},
that amplifies the small distance effect and hence the  size of the $1S$
amplitude relative to the $2S$ amplitude. As we will see, a factor in reducing $\mathcal{R} $ in the
molecular picture relative to the compact tetraquark is
the smaller size of the heavy-light systems, that is, that of  the $D^{(*)}$ mesons in the
molecular case, relative to the size of the diquark in the compact
tetraquark picture. Lastly, we will
see that in the molecular picture  the ratio $\mathcal{R}$ depends
quite sensitively on the size of the $1S$ and $2S$ charmonia. 

For this exploration we model $\Psi(r)$ in Eq.~\eqref{main}
% the BO
%wavefunction $\Psi_{\rm BO}$ in the molecular picture 
by the function in Eq.~\eqref{mol} for the molecule, while that of the compact tetraquark
by a Single Harmonic Oscillator\footnote{This is a ground state SHO oscillator wavefunction, chosen
  to reflect the linear potential between diquarks.} wave function:
$\Psi_{\rm SHO}(r)\sim e^{-\tfrac12 r^2/b^2} $. Both of these are functions of a
single parameter, the r.m.s. size of the state,
$R_{\rm rms}= \sqrt{\langle r^2\rangle}$.  In addition, $\chi_{_\text{M}}(r)$
and $\chi_{_\text{C}}(r)$ are both modeled by $\Psi_{\rm SHO}$~\footnote{This SHO function
  with $r_{\rm rms}=0.62\;\text{fm}$ is precisely the $D^{(*)}$ meson
  wavefunction in the ISGW model, which we have adopted for the
  molecular picture with an updated
  value, $r_{\rm rms}=0.69\;\text{fm}$. ISGW uses a linear combination of ground and
  first excited SHO functions in a variational principle with a
  Coulomb plus linear potential, and finds that the first excited
  state component is negligible.  The diquark is formed in a weaker
  potential, due to a reduction  in the string tension $\kappa \to
  \tfrac14\ \kappa$ in Eq.~\eqref{stension}, following the observation
  in footnote~\ref{scalcas},   resulting in a larger rms radius.}, each
characterized by a radius,
$r_{\rm rms}=\sqrt{\langle r^2\rangle}$  (we denote the size parameters as: $(R_{\text{rms}})_{_\text{M}}$ and $(r_{\text{rms}})_{D,D^*}$ for the molecule, and $(R_{\text{rms}})_{_\text{C}}$ and $(r_{\text{rms}})_{cu}$ for the compact tetraquark).

For the $\psi(1S)$ and $\psi (2S)$ wavefunctions 
we use SHO ground and first excited wavefunctions. With $b=0.4\;\text{fm}$ ($r_{\rm rms}=0.49\;\text{fm}$)
these are a good approximation to the numerical solutions 
 shown in
Fig.~\ref{reduced} that  are solutions of the Coulomb plus linear
potential.

\begin{figure}[t]
  \centering
  \includegraphics[width=0.40\textwidth]{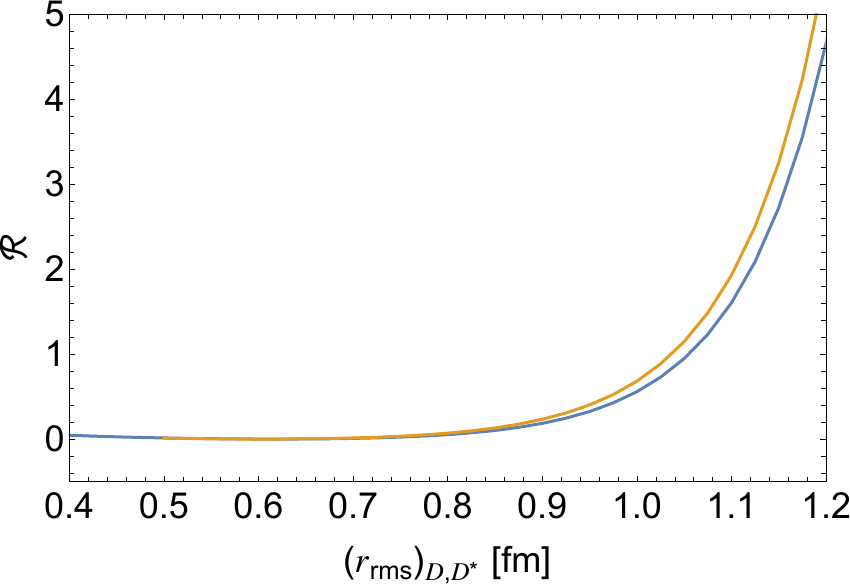}
   \caption{\footnotesize{  The ratio $\mathcal{R}$ of radiative decay branching
     fractions as a function of $D^{(*)}$ meson size, computed with
     the universal wave function $\Psi_{\rm mol.}(r)\sim\exp(- br)/r$ }(higher curve) and with a fake
     molecular function  $\Psi(r)\sim\exp(-b^2 r^2)/r$ (lower curve). In both cases the hadron size
     is fixed at $(R_{\rm rms})_{_\text{M}}=7\; \text{fm}$. The $\psi(1S)$ and $\psi (2S)$
     wavefunctions are approximated as ground and first excited state
     solutions to Schrodinger's SHO equation, with $r_{\rm
       rms}=0.5\;\text{fm}$ for the $1S$ state.   }
  \label{fig:fakeMolvsr}
\end{figure}

To see that the short distance singular structure of the universal
wavefunction in the molecular picture is responsible for the small
value of the ratio $\mathcal{R}$, we consider a modified
$\Psi$ of the form $e^{-\frac12 b^2 r^2}/r$, that is, an SHO ground
state divided by $r$ to amplify the short distance effects. In the
absence of the $1/r$ factor this is the wavefunction $\Psi_{\rm SHO}$ of our simplified
model for the compact tetraquark. 

Fig.~\ref{fig:fakeMolvsr} shows in blue the resulting ratio
$\mathcal{R}$ of this modified wavefunction  of the universal
wavefunction and in orange the result of the actual universal
wavefunction, both as function of the size of the %hadron.
molecule. The curves are remarkably similar, and although 
quantitatively different, they are both numerically much smaller than
the measured value as long as  $(r_{\rm rms})_{D,D^*}\lesssim1.1\;\text{fm}$.

\begin{figure}[ht!]
  \centering
  \includegraphics[width=0.45\textwidth]{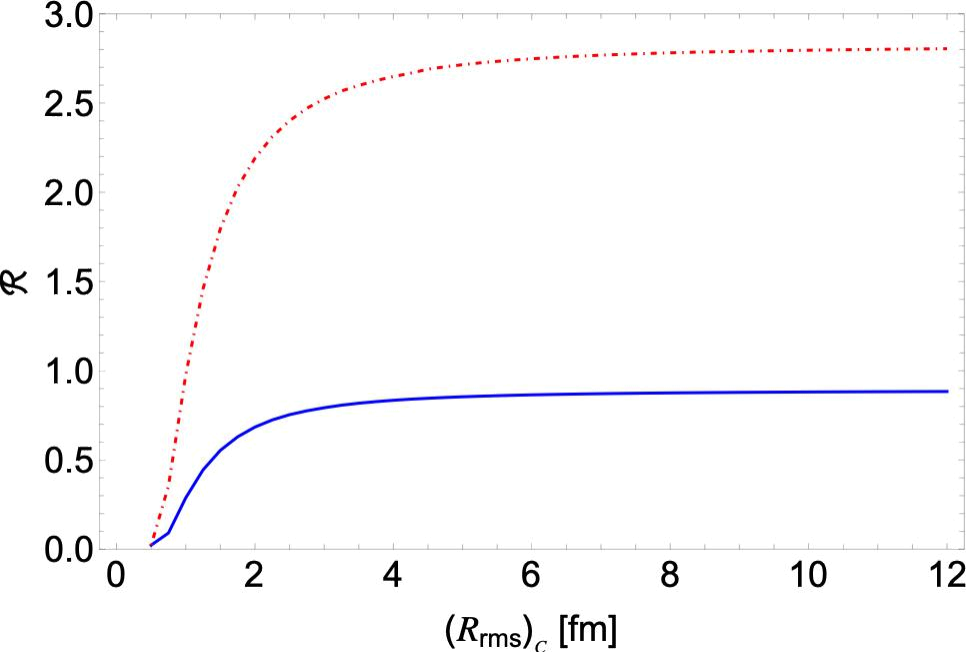}
  \caption{\footnotesize
 The ratio ${\cal R}$ as a function of the size $(R_{\text{rms}})_{_\text{C}}$ in the mock model of the compact tetraquark of Sec. 5, assuming $(r_{\text{rms}})_{cu} = (r_{\text{rms}})_{D,D^*} = 0.69$ fm (continuous curve)  or $(r_{\text{{rms}}})_{cu} =0.83$ fm (dot-dashed curve). 
% The $1S$ and $2S$ wavefunctions are as in Fig. 6. The difference of the two cases is expected since the binding force in the diquark is weaker than that in a singlet meson and we might expect a larger size for the diquark than the meson radius of the ISGW model. Correspondingly, a larger value of ${\cal R}$ results. 
      }
     \label{sizeComRrms}
  \end{figure}

Additional evidence that emphasis on  the small distance weight of the
overlap of wavefunctions produces a very suppressed ratio
$\mathcal{R}$ can be seen in Fig.~\ref{sizeComRrms}, which shows $\mathcal{R}$ in
the mock compact tetraquark model as a function of hadron size.  As the
wavefunction support concentrates around the origin, the contribution to
the $1S$ amplitude is accentuated while the one to the $2S$ amplitude is suppressed for small sizes of the $cu$ diquark, like those of $D, D^*$ mesons.

Turning to the other parametric dependence, we have already shown that the
ratio $\mathcal{R}$ depends quite sensitively on the size
$r_{\rm rms}$ of the $\chi_{_\text{M}}$ and $\chi_{_\text{C}}$ wavefunctions; see
Figs.~\ref{sizedq} and \ref{bgmol}.  We see that for both models the
ratio $\mathcal{R}$ increases rapidly with $(r_{\text{rms}})_{cu}$ for
$(r_{\text{rms}})_{cu}\gtrsim0.6\;\text{fm}$, and the molecular model exhibits a
minimum in the vicinity of
$(r_{\rm rms})_{D,D^*}\sim 0.55\;\text{fm}$. This minimum is at
  $\mathcal{R}=0$: it reflects the vanishing of the $2S$ amplitude at
  this particular meson size.

\begin{figure}[t]
  \centering
  \includegraphics[width=0.43\textwidth]{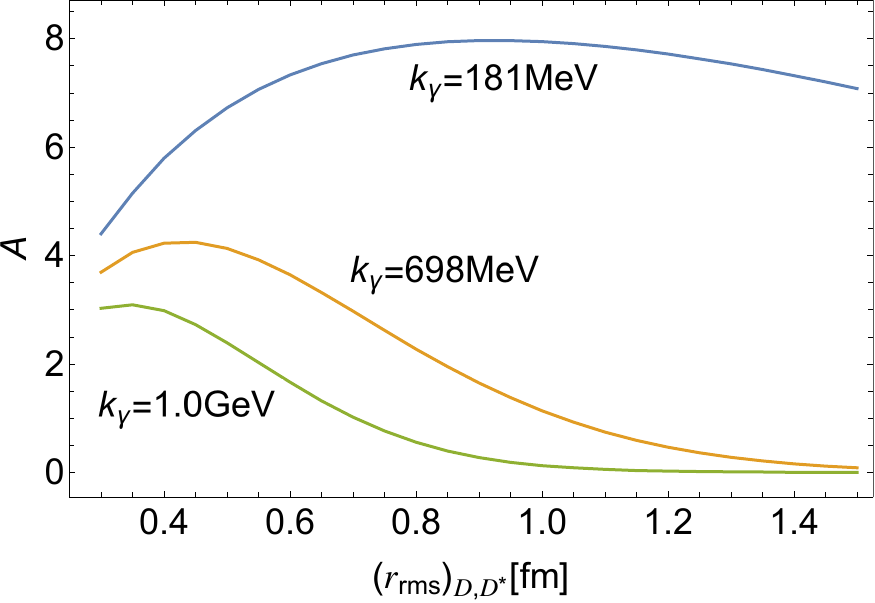}
    \includegraphics[width=0.45\textwidth]{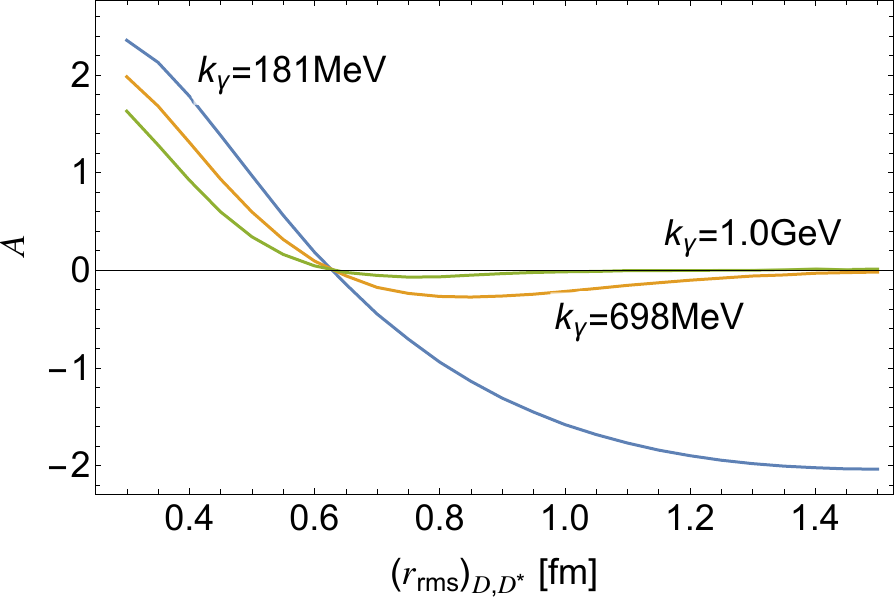}
  \caption[Photon energy dependence of $1S $ and $2S$ amplitudes]{\footnotesize{Amplitude for $X$
    radiative decay to $\psi(1S)$ (left panel) and  $\psi(2S)$ (right panel) as function of $D$-meson or diquark
    size, for three different values of (artificially modified) photon
    energy. The curves, in blue, orange and green,
     have $k=0.181\;\text{GeV}$ (as in the decay to $\psi(2S)$),
    $0.698\;\text{GeV}$ (as in the decay to $\psi(1S)$),  and $1.0\;\text{GeV}$, respectively.  This is
    computed at hadron size $R_{\rm rms}=7 \;\text{fm}$ with the
    approximate charmonium wavefunctions of Sec.~\ref{discussion1}  at
    $r_{\rm rms}=0.5\;\text{fm}$. }}
  \label{fig:amp-k-dep}
\end{figure}

  The rapid growth of these curves can be understood by comparing the $1S$
and $2S$ amplitudes as functions of $r_{\rm rms}$. They both decrease
towards zero, but the $1S$ amplitude does faster than the $2S$
amplitude. And the reason for this is that the oscillatory cosine
factor\footnote{The cosine factor is $$ 
\cos\left[k \left(\cos \lambda \left(\frac{R}{2}-r \cos \theta \right)-r \sin \theta  \sin \lambda  \cos \phi \right)\right]
 $$} in $\eqref{main}$ has shorter wavelength for $1S$ than for
$2S$. This can be verified by artificially changing the photon
energy~$k$ in Eq.~\eqref{main}.  Fig.~\ref{fig:amp-k-dep} shows the amplitude for $X$ radiative
decay to $\psi(1S)$ (left panel) and $\psi(2S)$ (right panel) as
function of $D$-meson or diquark size, for three different values of
(artificially modified) photon energy:
$k=0.181~\text{GeV}$ (blue), corresponding to the physical one for the decay
into $\psi(2S)$,  $0.698\;\text{GeV}$ (orange),
corresponding to the physical value for decay to $\psi(1S)$, and
$1.0\;\text{GeV}$ (green) corresponding to a lighter than
physical final state $\psi(1S)$. 
In the region $ (r_{\rm rms})_{D,D^*}\gtrsim0.5\;\text{fm}$ the physical $1S$
amplitude decreases, becoming negligible for  $ (r_{\rm rms})_{D,D^*}\gtrsim1.5\;\text{fm}$. 
At this meson size the physical $2S$
amplitude, while also decreasing,  is very sizable, leading to the
very enhanced value of  $\mathcal{R}$. The advertised zero in the
$2S$ amplitude is also made evident by these graphs.

\begin{figure}[t]
  \centering
 \includegraphics[width=0.45\textwidth]{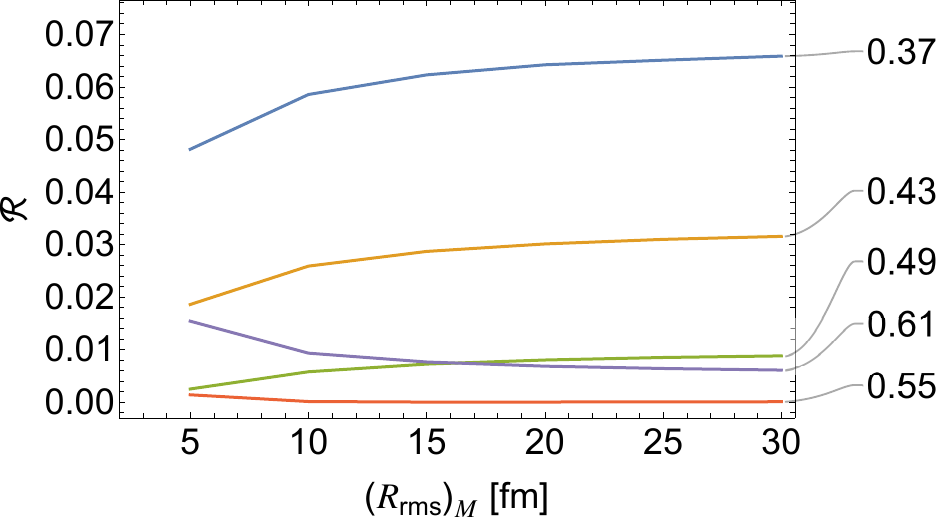} \hfill
 \includegraphics[width=0.45\textwidth]{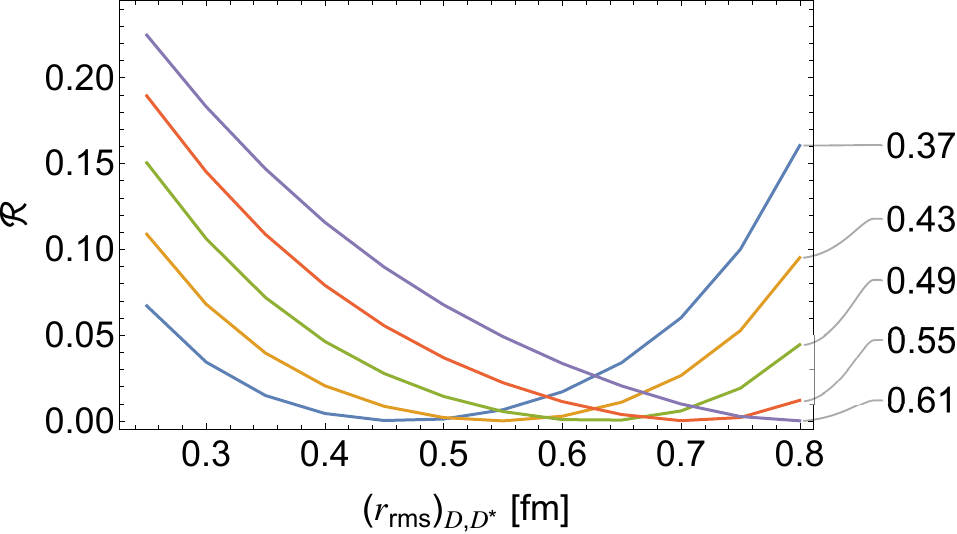} 
  \caption[$\mathcal{R}$ vs $R_{\rm rms}$(left) and
  $r_{\rm rms}$(right) for varying charmonium sizes]{\footnotesize{Ratio
    $\mathcal{R}$ of $2S$ to $1S$ branching fractions in the molecular
    model as a function of molecular size $(R_{\rm rms})_{_\text{M}}$ (left) and
    $D^{(*)}$-meson size $(r_{\rm rms})_{D,D^*}$ (right), for several artificially
    modified sizes of the $1S$ and $2S$ states. The curves are
    labeled by the rms size (fm) of the $1S$ approximate wavefunction
    of Sec.~\ref{discussion1}. The left plot is computed at meson size
  $(r_{\rm rms})_{D,D^*}=0.69\;\text{fm}$ while the right panel has hadron size
  $(R_{\rm rms})_{_\text{M}}=7\;\text{fm}$.}}
  \label{fig:many_onia}
\end{figure}

Lastly, we have investigated the dependence of $\mathcal{R}$ in the
molecular picture on the size of the final state charmonium states.
Fig.~\ref{fig:many_onia} shows the ratio $\mathcal{R}$ as a function
of molecular size $(R_{\rm rms})_{_\text{M}}$ (left) and $D^{(*)}$-meson size
$(r_{\rm rms})_{D,D^*}$, using the approximate wavefunctions described above,
for several artificial sizes of the $1S$ and $2S$ states. The left
plot, which  is computed at meson size
  $r_{\rm rms}=0.69\;\text{fm}$, shows that $\mathcal{R}$ remains very
  small and fairly constant as a function of hadron size. The right panel has molecular size
  $(R_{\rm rms})_{_\text{M}}=10\;\text{fm}$, although its precise value is
  irrelevant, as shown by the left panel. One sees that as the size of
  the charmonium states decreases, the location of the minimum in these curves 
  moves left, towards a smaller value of $(r_{\rm rms})_{D,D^*}$. Any
  attempt to increase $\mathcal{R}$ in the molecular model by
  decreasing the size of charmonium states by changing the interquark
  potential would be frustrated by a corresponding decrease in the
  size of the $D$-meson.\footnote{Unless one invokes different quark
    forces in the heavy-light system from  the heavy-heavy one. }

\section{Conclusions}
We have computed the ratio of branching fractions ${\cal R}=
{\cal B}(X\to \psi^\prime \gamma)/{\cal B}(X\to \psi \gamma)$ in the
molecular hypothesis following~\cite{Voloshin:2003nt,
  Braaten:2003he}%and~\cite{Isgur:1988gb}
, finding  ${\cal R}\approx0.04$, a value
much smaller than ${\cal R}_{\rm exp}\simeq 6\pm
4$\cite{ParticleDataGroup:2022pth}. While we assume 
binding energies around $B\sim O(100)$~KeV, leading to molecular sizes
of $\sim10\;\text{fm}$, we show that the result is quite 
insensitive to the characteristic   molecular size. In contrast,
$\mathcal{R}$ is a fast increasing functions of the size of the
$D^{(*)}$ meson, and our result is obtained for 0.69\;fm. 
The price to pay to  get larger ${\cal R}$ values in the molecular
hypothesis is that of allowing larger  $D$
and $D^*$ mesons,  well above expectations.

We present a similar calculation 
for the $X$ described as a compact tetraquark, that  is, a diquark-antidiquark  color-molecule,
treated in the Born-Oppenheimer
approximation\cite{Maiani:2019cwl}. We find that the result is
 also  very sensitive to the size of the diquark and not much to the size
 of the whole tetraquark.
 This time however, sizes in excess of 0.7 fm are expected, 1\;fm not unrealistic.
  Using the parameters for
  the compact tetraquark in Ref.\cite{Maiani:2019cwl} we
  find  ${\cal R}\sim 0.95$, about thirty times
  larger than the predicted value in the molecular model. 
  This value corresponds to the most conservative determination we
  have (with rms radius $\simeq 0.8$~fm). 
  If we adopt  the Casimir scaling of the string tension used to determine the 
 diquark orbitals described in Section~\ref{compa}, and therefore
 allow slightly larger rms radius, 
  we obtain significantly  larger ${\cal R}$ values as in  Fig.~\ref{sizedq}.

 The conclusions obtained are rather robust under
  reasonable parameter variation and even changes in wavefunction
  shape. Hence,  we have found, rather remarkably, that there is a qualitative
  distinction between what is obtained in the molecular and compact
  pictures 
    
    A better knowledge of the experimental uncertainty in ${\cal R}_{\rm
    exp}$ would be extremely helpful in clarifying  the true nature of the $X(3872)$

\section*{Acknowledgements}
We wish to thank Vanya Belyaev for interesting discussions on this topic. 
The work of B.G. was supported in part by the Sapienza University visiting
professor program and by the U.S. Department of Energy Grant No. DE-SC0009919.  

\bibliographystyle{unsrt}

\end{document}